\documentclass[prd,aps,nofootinbib,floatfix,10pt]{revtex4}

\usepackage{amsmath,graphicx,epsfig,amssymb,dsfont,mathtools}
\usepackage[usenames]{color}
\usepackage{ulem} %% for strike-through
\usepackage{bigstrut}
\usepackage{slashed}

\begin{document}
\title{$B_c\to B_{sJ}$ form factors and $B_c$ decays into $B_{sJ}$ in   covariant light-front approach}
\author{Yu-Ji Shi$^1$~\footnote{Email:shiyuji@sjtu.edu.cn}, Wei Wang $^{1,2}$~\footnote{Email:wei.wang@sjtu.edu.cn}, Zhen-Xing Zhao$^1$~\footnote{Email:star\_0027@sjtu.edu.cn}}
\affiliation{
$^1$ INPAC, Shanghai Key Laboratory for Particle Physics and Cosmology, Department of Physics and Astronomy, Shanghai Jiao-Tong University, Shanghai, 200240,   China\\
$^2$ State Key Laboratory of Theoretical Physics, Institute of Theoretical Physics, Chinese Academy of Sciences, Beijing 100190, China
 }

\begin{abstract}
We suggest to study the $B_{s}$ and its excitations   $B_{sJ}$ in the $B_c$ decays.
We calculate the $B_c\to B_{sJ}$ and $B_c\to B_{J}$ form factors within the covariant light-front quark model, where the $B_{sJ}$ and $B_{J}$ denote an s-wave or p-wave $\bar bs$ and $\bar bd$ meson, respectively.  The form factors  at $q^2=0$ are directly computed  while their $q^2$-distributions are obtained by the extrapolation.  The derived form factors are then used to study semileptonic $B_c\to (B_{sJ},B_{J})\bar\ell\nu$  decays,  and   nonleptonic $B_c\to B_{sJ}\pi$. Branching fractions and polarizations are  predicted in the standard model. We find that the   branching fractions are sizable and might be accessible  at the LHC experiment and future high-energy $e^+e^-$ colliders with a high luminosity at the $Z$-pole.  The future experimental measurements are helpful to study the nonperturbative QCD dynamics  in the presence of a heavy spectator and also of great value for the spectroscopy study.
\end{abstract}

\maketitle

\section{Introduction}

In the past decades, there have been  a lot of progresses  in hadron spectroscopy, thanks to the well-operating   experiments including the $e^+e^-$ colliders and hadron colliders.  The immense  interest in    spectroscopy    is not only due to the fact that one is  able to find many missing   hadrons to complete  the quark model, but more importantly   due to the observations of states that are  unexpected  in the simple quark model.   The latter ones are generally called   hadron exotics.   A milestone  in the exotics exploration  is the discovery of $X(3872)$, firstly in $B$ decays by Belle Collaboration~\cite{Choi:2003ue} and subsequently   confirmed in many distinct processes  in different experiments~\cite{Acosta:2003zx,Aubert:2004ns,Abazov:2004kp}.  It was found the properties of this meson is peculiar.  Since then  the identification of   multiquark hadrons becomes  a hot  topic in hadron physics.  Inspired by the discovery of $X(3872)$,  a number of new interesting structures were discovered in the mass region of heavy quarkonium.  Refer to Refs.~\cite{Brambilla:2010cs,Esposito:2014rxa,Chen:2016qju,Ali:2016gli} for   recent reviews.

On theoretical side,  deciphering the underlying dynamics of these multiquark states is a formidable challenge, and is  often   based on explicit and  distinct assumptions. In many  assumptions, the quarkonium-like states are usually  composed  of a pair of heavy constituents, which makes it vital to study first  the heavy-light hadron.
In the system with one    heavy charm quark, a series of important results  start with the discoveries of the narrow states $D_s(2317)$ in the $D_s^+\pi^0$
final state and $D_s(2460)$ in the $D_s^*\pi^0$ and $D_s\gamma$
final state~\cite{Aubert:2003fg,Besson:2003cp}. Along this line, a  few other new states, such as $D_{s1}(2536),D_{s2}(2573),D_s(2710)$, have been observed
at the B factory and other facilities~\cite{Agashe:2014kda}.

%Towards the understanding  of these newly observed states and exploration
%of the different assignments, the study of  the bottom-counterparts $B_{sJ}$ states  is one of the most
%ideal platforms, since they are related by the heavy quark symmetry.
Bottomed hadrons are related to charmed mesons by   heavy quark symmetry.
But compared to the charm sector, there are less progresses in the bottomed hadrons.
In experiment, only a few bottom-strange mesons are observed, and most of which are believed to be filled in the quark model.  In this paper, we propose to use the $B_c$ decays  and study the spectrum of the $B_{sJ}$. It gains   a few advantages.
First, the large production rates of the $B_c$ is in expectation,  in particular the LHCb will produce a number of
$B_c$ events and thus the $B_c\to B_{sJ}$ decays will have a large
potential to be observed. Secondly, the scale over the $m_W$ can be
computed in the perturbation theory and the QCD evolution between
the $m_W$ and the low energy scale $m_c$ is well organized by making
use of the renormalization group improved perturbation theory.
Consequently, the   $B_c$ decays into $B_s$  and other excited states have  received some theoretical
attentions~\cite{Chang:1992pt,Lu:1995ug,Liu:1997hr,Colangelo:1999zn,Ivanov:2000aj,Kiselev:2000pp,AbdElHady:1999xh,Choudhury:2001ty,Ebert:2003wc,Sun:2008wa,Ebert:2010zu,Fu:2011tn,Naimuddin:2012dy,Xiao:2013lia,Sun:2014jka,Sun:2014ika,Sun:2015exa,He:2016xvd,Choudhury:2016rgp}.
In the following  we will be dedicated to investigate the production
rates of $B_{sJ}$  meson (an s-wave or p-wave $\bar bs$ hadron) in semileptonic and nonleptonic $B_c$
mesons decays under the framework of the covariant light-front quark
model (LFQM)~\cite{Jaus:1999zv}.

In the $B_c\to B_{sJ}$ decays, the quark level transition is the $c\to s$  in which the heavy bottom quark acts as a spectator. Since most of the momentum of the hadron is carried  by the spectator, there is no large momentum transfer and   the transition is dominated by the soft mechanism.  A form factor can then be expressed as a overlap of the wave functions of the initial and final state hadrons. Treatments in quark models like the LFQM are of this type.

As pointed out
in Ref.~\cite{Brodsky:1997de}, the light front   approach owns some unique
features which are   suitable to handle  a hadronic
bound state.  The LFQM~\cite{Jaus:1989au,Jaus:1991cy,Cheng:1996if,Choi:2001hg}
can provide a relativistic treatment of moving hadrons
and   give a fully treatment of  hadron spins in terms of the
Melosh rotation.  Light-front wave functions, which
characterize  the hadron in terms of their fundamental quark and gluon
degrees of freedom, are independent of  hadron momentum and
thus are  Lorentz invariant. Moreover, in covariant
LFQM~\cite{Jaus:1999zv}, the spurious contribution which
depends on the orientation of  light-front is elegantly  eliminated
by including  zero-mode contributions. This covariant
model has been successfully extended to study the decay constants
and form factors of various mesons~\cite{Cheng:2003sm,Cheng:2004yj,Ke:2009ed,Ke:2009mn,Cheng:2009ms,Lu:2007sg,Wang:2007sxa,Wang:2008xt,Wang:2008ci,Wang:2009mi,Chen:2009qk,Li:2010bb,Verma:2011yw}.
Through this study of $B_c\to B_{sJ}$ in LFQM, we believe that one will not only gain the
information about the  decay dynamics in the presence of a heavy spectator  but will also
provide a side-check for the classification of the  heavy-light  mesons. It is also helpful   towards the  establishment of  a global  picture  of the heavy-light spectroscopy including the exotic  spectrum.

The rest of this paper is organized as follows.  In Sec.~\ref{sec:FF-CLFQM}, we will give a brief description of  the parametrization of form factors,  the framework of covariant LFQM, and the form factor calculation in this model.   We present our numerical results  for various transitions in Sec.~\ref{sec:numerics}.  In Sec.~\ref{sec:application}, we use the form factors to study semileptonic and nonleptonic $B_c$ decays. In this section, we will present our  predictions for    branching fractions and polarizations. The last section contains a brief summary.

%%%%%%%%%%%%%%%%%%%%%%%%%%%%%%%%%%%%%%%%%%%%%%%%%%%%%%%%%%%%%%%%%
%%%%%%%%%%%%%%%%%%%%%%%%%%%%%%%%%%%%%%%%%%%%%%%%%%%%%%%%%%%%%%%%%
\section{Transition form factors in the covariant LFQM}
\label{sec:FF-CLFQM}
%%%%%%%%%%%%%%%%%%%%%%%%%%%%%%%%%%%%%%%%%%%%%%%%%%%%%%%%%%%%%%%%%
%%%%%%%%%%%%%%%%%%%%%%%%%%%%%%%%%%%%%%%%%%%%%%%%%%%%%%%%%%%%%%%%%

\subsection{$B_{c}\to B_{sJ}$ form factors}

The effective electroweak Hamiltonian  for the $B_{c}\to B_{sJ}\bar{l}\nu$
reads
\begin{equation}
{\cal H}_{{\rm eff}}=\frac{G_{F}}{\sqrt{2}}V_{cs}^{*}[\bar{s}\gamma_{\mu}(1-\gamma_{5})c][\bar{\nu}\gamma^{\mu}(1-\gamma_{5})l],
\end{equation}
where the $G_{F}$ and $V_{cs}$ is   Fermi constant and   Cabibbo-Kobayashi-Maskawa
matrix element, respectively. Leptonic parts can be computed in  perturbation theory while hadronic contributions are paraemtrized in terms of form factors.

An $s$-wave meson corresponds to a pseudo-scalar meson or a vector meson, abbreviated as $P$ and $V$ respectively. For a $p$-wave meson, the involved state is a scalar $S$, an axial-vector $A$ or a tensor meson $T$.
In the following we introduce the abbreviations  $P=P^{\prime}+P^{\prime\prime}$, $q=P^{\prime}-P^{\prime\prime}$
and adopt the convention of $\epsilon_{0123}=1$.  The $B_{c}\to P,V$ form
factors can be defined as follows: 
\begin{eqnarray}
\langle P(P^{\prime\prime})|V_{\mu}|B_{c}(P^{\prime})\rangle & = & \left(P_{\mu}-\frac{m_{B_{c}}^{2}-m_{P}^{2}}{q^{2}}q_{\mu}\right)F_{1}^{B_{c}P}(q^{2}) +\frac{m_{B_{c}}^{2}-m_{P}^{2}}{q^{2}}q_{\mu}F_{0}^{B_{c}P}(q^{2}),\nonumber \\
\langle V(P^{\prime\prime},\varepsilon^{\prime\prime})|V_{\mu}|B_{c}(P^{\prime})\rangle & = & -\frac{1}{m_{B_{c}}+m_{V}}\,\epsilon_{\mu\nu\alpha\beta}\varepsilon^{\prime\prime}{}^{*\nu}P^{\alpha}q^{\beta}V^{B_{c}V}(q^{2}),\nonumber \\
\langle V(P^{\prime\prime},\varepsilon^{\prime\prime})|A_{\mu}|B_{c}(P^{\prime})\rangle & = & 2i m_{V}\,\frac{\varepsilon^{\prime\prime*}\cdot q}{q^{2}}\,q_{\mu}A_{0}^{B_{c}V}(q^{2})+i(m_{B_{c}}+m_{V})A_{1}^{B_{c}V}(q^{2})\left[\varepsilon_{\mu}^{\prime\prime*}  - \frac{\varepsilon^{\prime\prime*}\cdot q}{q^2}q_\mu\right] \nonumber \\
 &  &  -i\frac{\varepsilon^{\prime\prime*}\cdot P}{m_{B_{c}}+m_{V}}\,A_{2}^{B_{c}V}(q^{2})\left[P_{\mu}- \frac{m_{B}^2-m_{V}^2}{q^2} q_\mu \right].
\end{eqnarray}
In analogy with $B_{c}\to V$ form factors, we parametrize
$B_{c}\to T$ form factors as
\begin{eqnarray}
\langle T(P^{\prime\prime},\varepsilon^{\prime\prime})|V_{\mu}|B_{c}(P^{\prime})\rangle & = & -\frac{2V^{B_{c}T}(q^{2})}{m_{B_{c}}+m_{T}}\epsilon^{\mu\nu\rho\sigma}(\varepsilon_{T}^{*})_{\nu}(P^{\prime})_{\rho}(P^{\prime\prime})_{\sigma},\nonumber \\
\langle T(P^{\prime\prime},\varepsilon^{\prime\prime})|A_{\mu}|B_{c}(P^{\prime})\rangle & = & 
2im_{T}\frac{\varepsilon_{T}^{*}\cdot q}{q^{2}}q_{\mu} A_{0}^{B_{c}T}(q^{2})+i(m_{B_{c}}+m_{T})A_{1}^{B_{c}T}(q^{2})\Bigg[\varepsilon_{T\mu}^{*} -\frac{\varepsilon_{T}^{*}\cdot q}{q^{2}}q_{\mu}\Bigg]\nonumber \\
 &  & -i\frac{\varepsilon_{T}^{*}\cdot q}{m_{B_{c}}+m_{T}}A_{2}^{B_{c}T}(q^{2})\left[P_{\mu}-\frac{m_{B_{c}}^{2}-m_{T}^{2}}{q^{2}}q_{\mu}\right],
\end{eqnarray}
with
\begin{equation}
\varepsilon_{T\mu}(h)=\frac{1}{m_{B_{c}}}\varepsilon_{\mu\nu}^{\prime\prime}(h)P^{\prime\nu}.
\end{equation}
The $B_{c}\to S,A$ form factors can be defined by exchanging the vector and axial-vector current:  
\begin{eqnarray}
\langle S(P^{\prime\prime})|A_{\mu}|B_{c}(P^{\prime})\rangle & = & -i\Bigg[\left(P_{\mu}-\frac{m_{B_{c}}^{2}-m_{S}^{2}}{q^{2}}q_{\mu}\right)F_{1}^{B_{c}S}(q^{2}) +\frac{m_{B_{c}}^{2}-m_{S}^{2}}{q^{2}}q_{\mu}F_{0}^{B_{c}S}(q^{2})\Bigg],\nonumber \\
\langle A(P^{\prime\prime},\varepsilon^{\prime\prime})|V_{\mu}|B_{c}(P^{\prime})\rangle & = &
-2m_{A}\,\frac{\varepsilon^{\prime\prime*}\cdot q}{q^{2}}\,q_{\mu}V_{0}^{B_{c}A}(q^{2})-(m_{B_{c}}+m_{A})V_{1}^{B_{c}A}(q^{2})\left[\varepsilon_{\mu}^{\prime\prime*}  - \frac{\varepsilon^{\prime\prime*}\cdot q}{q^2}q_\mu\right] \nonumber \\
 &  &  +\frac{\varepsilon^{\prime\prime*}\cdot P}{m_{B_{c}}+m_{A}}\,V_{2}^{B_{c}A}(q^{2})\left[P_{\mu}- \frac{m_{B}^2-m_{A}^2}{q^2} q_\mu \right],\nonumber \\
\langle A(P^{\prime\prime},\varepsilon^{\prime\prime})|A_{\mu}|B_{c}(P^{\prime})\rangle & = & -i\frac{1}{m_{B_{c}}-m_{A}}\,\epsilon_{\mu\nu\alpha\beta}\varepsilon^{\prime\prime}{}^{*\nu}P^{\alpha}q^{\beta}A^{B_{c}A}(q^{2}).
\end{eqnarray}

The spin-2 polarization tensor can be
constructed using the standard polarization vector $\varepsilon$:
\begin{eqnarray}
 &  & \varepsilon_{\mu\nu}^{\prime\prime}(P^{\prime\prime},\pm2)=\varepsilon_{\mu}(\pm)\varepsilon_{\nu}(\pm),\;\;\;\;\varepsilon_{\mu\nu}^{\prime\prime}(P^{\prime\prime},\pm1)=\frac{1}{\sqrt{2}}[\varepsilon_{\mu}(\pm)\varepsilon_{\nu}(0)+\varepsilon_{\nu}(\pm)\varepsilon_{\mu}(0)],\nonumber \\
 &  & \varepsilon_{\mu\nu}^{\prime\prime}(P^{\prime\prime},0)=\frac{1}{\sqrt{6}}[\varepsilon_{\mu}(+)\varepsilon_{\nu}(-)+\varepsilon_{\nu}(+)\varepsilon_{\mu}(-)]+\sqrt{\frac{2}{3}}\varepsilon_{\mu}(0)\varepsilon_{\nu}(0).
\end{eqnarray}
It is  symmetric and traceless, and $\varepsilon_{\mu\nu}^{\prime\prime}P^{\prime\prime\nu}=0$.
If the recoiling meson is moving on the plus direction of the $z$
axis, their explicit structures are chosen as 
\begin{eqnarray}
\varepsilon_{\mu}(0) & = & \frac{1}{m_{T}}(|\vec{p}_{T}|,0,0,-E_{T}),\;\;\;
\varepsilon_{\mu}(\pm)   =   \frac{1}{\sqrt{2}}(0,\pm1,i,0),
\end{eqnarray}
where $E_{T}$ and $|\vec{p}_{T}|$ are the energy and the momentum magnitude
of the tensor meson  in the $B_{c}$   rest frame, respectively.

%%%%%%%%%%%%%%%%%%%%%%%%%%%%%%%%%%
\subsection{Covariant light-front approach}
%%%%%%%%%%%%%%%%%%%%%%%%%%%%%%%%%%

In the covariant LFQM, it is convenient to use the light-front decomposition
of the momentum $P^{\prime}=(P^{\prime-},P^{\prime+},P_{\perp}^{\prime})$,
with $P^{\prime\pm}=P^{\prime0}\pm P^{\prime3}$, and thus $P^{\prime2}=P^{\prime+}P^{\prime-}-P_{\perp}^{\prime2}$.
The incoming (outgoing) meson has the momentum  $P^{\prime}=p_{1}^{\prime}+p_{2}$
($P^{\prime\prime}=p_{1}^{\prime\prime}+p_{2}$) and the mass  $M^{\prime}$
($M^{\prime\prime}$). The quark and antiquark inside the incoming
(outgoing) meson have the mass $m_{1}^{\prime(\prime\prime)}$ and
$m_{2}$, respectively.  Their momenta are denoted as $p_{1}^{\prime(\prime\prime)}$
and $p_{2}$ respectively. In particular these momenta can be written
in terms of the internal variables $(x_{i},p_{\perp}^{\prime})$ by:
\begin{equation}
p_{1,2}^{\prime+}=x_{1,2}P^{\prime+},\quad p_{1,2\perp}^{\prime}=x_{1,2}P_{\perp}^{\prime}\pm p_{\perp}^{\prime},
\end{equation}
with the momentum fractions $x_{1}+x_{2}=1$. With these internal variables, one can define
some useful quantities for both incoming and outgoing mesons:
\begin{eqnarray}
M_{0}^{\prime2} & = & (e_{1}^{\prime}+e_{2})^{2}=\frac{p_{\perp}^{\prime2}+m_{1}^{\prime2}}{x_{1}}+\frac{p_{\perp}^{\prime2}+m_{2}^{2}}{x_{2}}, \;\; \widetilde{M}_{0}^{\prime}  =  \sqrt{M_{0}^{\prime2}-(m_{1}^{\prime}-m_{2})^{2}},\nonumber \\
e_{i}^{(\prime)} & = & \sqrt{m_{i}^{(\prime)2}+p_{\perp}^{\prime2}+p_{z}^{\prime2}},\; \;\;
p_{z}^{\prime}  =  \frac{x_{2}M_{0}^{\prime}}{2}-\frac{m_{2}^{2}+p_{\perp}^{\prime2}}{2x_{2}M_{0}^{\prime}}.
\end{eqnarray}
%where  the $e_{i}^{(\prime)}$ can be treated as  the energy of the quark $i$.

%%%%%%%%%%%%%%%%%%%%%%%%%%%%%%%
\begin{table}
\caption{Meson-quark-antiquark vertices used in the covariant LFQM. In the
case of the outgoing meson, one should use instead $i(\gamma_{0}\Gamma_{M}^{\prime\dagger}\gamma_{0})$
for the corresponding vertices.}
\label{Tab:vertices}
\begin{center}
\begin{tabular}{|c|c|}
\hline
$M(\,^{2S+1}L_{J})$ & $i\Gamma_{M}^{\prime}$\tabularnewline
\hline
\hline
pseudoscalar ($\,^{1}S_{0}$) & $H_{P}^{\prime}\gamma_{5}$\tabularnewline
\hline
scalar ($\,^{3}P_{0}$) & $H_{S}^{\prime}$\tabularnewline
\hline
vector ($\,^{3}S_{1}$) & $iH_{V}^{\prime}[\gamma_{\mu}-\frac{1}{W_{V}^{\prime}}(p_{1}^{\prime}-p_{2})_{\mu}]$\tabularnewline
\hline
axial ($\,^{3}P_{1}$) & $iH_{^{3}A}^{\prime}[\gamma_{\mu}+\frac{1}{W_{^{3}A}^{\prime}}(p_{1}^{\prime}-p_{2})_{\mu}]\gamma_{5}$\tabularnewline
\hline
axial ($\,^{1}P_{1}$) & $iH_{^{1}A}^{\prime}[\frac{1}{W_{^{1}A}^{\prime}}(p_{1}^{\prime}-p_{2})_{\mu}]\gamma_{5}$\tabularnewline
\hline
tensor ($\,^{3}P_{2}$) & $i\frac{1}{2}H_{T}^{\prime}[\gamma_{\mu}-\frac{1}{W_{T}^{\prime}}(p_{1}^{\prime}-p_{2})_{\mu}](p_{1}^{\prime}-p_{2})_{\nu}$\tabularnewline
\hline
\end{tabular}
\end{center}
\end{table}
%%%%%%%%%%%%%%%%%%%%%%%%%%%%%%%

%%%%%%%%%%%%%%%%%%%%%%%%%%%%%%%%%%%%%%%%%%%%%%%%%%%%%
\begin{figure}
\includegraphics[scale=0.5]{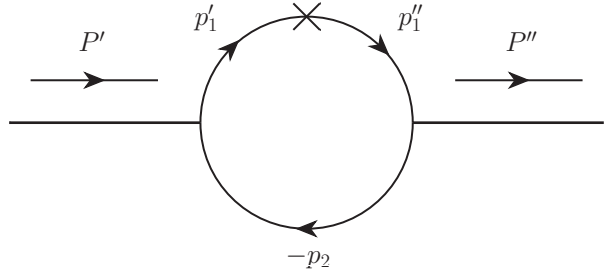}
\caption{Feynman diagram for transition form factors, where the
cross symbol in the diagram denotes the  transition
current.   }
\label{fig:feyn}
\end{figure}
%%%%%%%%%%%%%%%%%%%%%%%%%%%%%%%%%%%%%%%%%%%%%%%%%%%%%

Feynman rules for meson-quark-antiquark vertices can be derived using
the conventional light-front approach, whose forms for the s-wave and p-wave states are collected  in Table~\ref{Tab:vertices}~\cite{Jaus:1999zv,Cheng:2003sm}.  An extension to  the d-wave vertices has been conducted in Ref.~\cite{Ke:2011mu}.  
In the following  we will  take  the $B_{c}\to B_{s}$ transition as the example and   illustrate the calculation.
To do so, we will  consider the matrix element
\begin{equation}
\langle P(P^{\prime\prime})|V_{\mu}|P(P^{\prime})\rangle\equiv{\cal B}_{\mu}^{PP},
\end{equation}
whose  Feynman diagram  is shown in Fig. \ref{fig:feyn}. It is straightforward to obtain
\begin{equation}
{\cal B}_{\mu}^{PP}=i^{3}\frac{N_{c}}{(2\pi)^{4}}\int d^{4}p_{1}^{\prime}\frac{H_{P}^{\prime}H_{P}^{\prime\prime}}{N_{1}^{\prime}N_{1}^{\prime\prime}N_{2}}S_{V\mu}^{PP},
\end{equation}
where
\begin{eqnarray}
S_{V\mu}^{PP} & = & {\rm Tr}[\gamma_{5}(\slashed p_{1}^{\prime\prime}+m_{1}^{\prime\prime})\gamma_{\mu}(\slashed p_{1}^{\prime}+m_{1}^{\prime})\gamma_{5}(\slashed p_{2}-m_{2})],\nonumber \\
N_{1}^{\prime(\prime\prime)} & = & p_{1}^{\prime(\prime\prime)2}-m_{1}^{\prime(\prime\prime)2}+i\epsilon,
\;\;\;
N_{2}   =   p_{2}^{2}-m_{2}^{2}+i\epsilon.
\end{eqnarray}
Here we consider the $q^{+}=0$ frame. The $p_{1}^{\prime-}$
integration picks up the pole $p_{2}=\hat{p}_{2}=[({p_{2\perp}^{2}+m_{2}^{2}})/{p_{2}^{+}},p_{2}^{+},p_{2\perp}]$
and leads to
\begin{eqnarray}
N_{1}^{\prime(\prime\prime)} & \to & \hat{N}_{1}^{\prime(\prime\prime)}=x_{1}(M^{\prime(\prime\prime)2}-M_{0}^{\prime(\prime\prime)2}),\nonumber \\
H_{P}^{\prime(\prime\prime)} & \to & h_{P}^{\prime(\prime\prime)},\nonumber \\
\int\frac{d^{4}p_{1}^{\prime}}{N_{1}^{\prime}N_{1}^{\prime\prime}N_{2}}H_{P}^{\prime}H_{P}^{\prime\prime}S^{PP} & \to & -i\pi\int\frac{dx_{2}d^{2}p_{\perp}^{\prime}}{x_{2}\hat{N}_{1}^{\prime}\hat{N}_{1}^{\prime\prime}}h_{P}^{\prime}h_{P}^{\prime\prime}\hat{S}^{PP},
\end{eqnarray}
where
\begin{equation}
M_{0}^{\prime\prime2}=\frac{p_{\perp}^{\prime\prime2}+m_{1}^{\prime\prime2}}{x_{1}}+\frac{p_{\perp}^{\prime\prime2}+m_{2}^{2}}{x_{2}},
\end{equation}
with $p_{\perp}^{\prime\prime}=p_{\perp}^{\prime}-x_{2}q_{\perp}$.
The explicit form  of $h_{P}^{\prime}$ has been derived in Ref.~\cite{Jaus:1999zv,Cheng:2003sm}
\begin{equation}
h_{P}^{\prime}=(M^{\prime2}-M_{0}^{\prime2})\sqrt{\frac{x_{1}x_{2}}{N_{c}}}\frac{1}{\sqrt{2}\widetilde{M}_{0}^{\prime}}\varphi^{\prime},
\end{equation}
where $\varphi^{\prime}$ is the light-front momentum distribution
amplitude for $s$-wave meson. In practice, the following Gaussian-type
wave function can be adopted \cite{Jaus:1999zv,Cheng:2003sm}:
\begin{equation}
\varphi^{\prime}=\varphi^{\prime}(x_{2},p_{\perp}^{\prime})=4\left(\frac{\pi}{\beta^{\prime2}}\right)^{3/4}\sqrt{\frac{dp_{z}^{\prime}}{dx_{2}}}\exp\left(-\frac{p_{z}^{\prime2}+p_{\perp}^{\prime2}}{2\beta^{\prime2}}\right).
\end{equation}

As shown in Ref.~\cite{Jaus:1999zv,Cheng:2003sm}, the inclusion of the so-called zero mode contribution
in the above matrix elements in practice amounts to the replacements
\begin{eqnarray*}
\hat{p}_{1\mu}^{\prime} & \doteq & P_{\mu}A_{1}^{(1)}+q_{\mu}A_{2}^{(1)}, \;\;\;
\hat{N}_{2}   \to  Z_{2}, \;\;\;
\hat{p}_{1\mu}^{\prime}\hat{N}_{2}  \to  q_{\mu}\left[A_{2}^{(1)}Z_{2}+\frac{q\cdot P}{q^{2}}A_{1}^{(2)}\right],
\end{eqnarray*}
where the symbol $\doteq$ in the above equation reminds us that it
is true only in the ${\cal B}_{\mu}^{PP}$ integration. $A_{j}^{(i)}$ and $Z_{2}$, which are functions of $x_{1,2}$, $p_{\perp}^{\prime2}$, $p_{\perp}^{\prime}\cdot q_{\perp}$
and $q^{2}$, are listed in Appendix \ref{app:FF}. After the replacements, we arrive at
\begin{eqnarray}
f_{+}(q^{2}) & = & \frac{N_{c}}{16\pi^{3}}\int dx_{2}d^{2}p_{\perp}^{\prime}\frac{h_{P}^{\prime}h_{P}^{\prime\prime}}{x_{2}\hat{N}_{1}^{\prime}\hat{N}_{1}^{\prime\prime}}\Bigg[x_{1}(M_{0}^{\prime2}+M_{0}^{\prime\prime2})+x_{2}q^{2}-x_{2}(m_{1}^{\prime}-m_{1}^{\prime\prime})^{2}-x_{1}(m_{1}^{\prime}-m_{2})^{2}-x_{1}(m_{1}^{\prime\prime}-m_{2})^{2}\Bigg],\nonumber \\
f_{-}(q^{2}) & = & \frac{N_{c}}{16\pi^{3}}\int dx_{2}d^{2}p_{\perp}^{\prime}\frac{2h_{P}^{\prime}h_{P}^{\prime\prime}}{x_{2}\hat{N}_{1}^{\prime}\hat{N}_{1}^{\prime\prime}}\Bigg\{-x_{1}x_{2}M^{\prime2}-p_{\perp}^{\prime2}-m_{1}^{\prime}m_{2}+(m_{1}^{\prime\prime}-m_{2})\nonumber \\
 &  & \times(x_{2}m_{1}^{\prime}+x_{1}m_{2})+2\frac{q\cdot P}{q^{2}}\left(p_{\perp}^{\prime2}+2\frac{(p_{\perp}^{\prime}\cdot q_{\perp})^{2}}{q^{2}}\right)+2\frac{(p_{\perp}^{\prime}\cdot q_{\perp})^{2}}{q^{2}}\nonumber \\
 &  & -\frac{p_{\perp}^{\prime}\cdot q_{\perp}}{q^{2}}\Bigg[M^{\prime\prime2}-x_{2}(q^{2}+q\cdot P)-(x_{2}-x_{1})M^{\prime2}+2x_{1}M_{0}^{\prime2} -2(m_{1}^{\prime}-m_{2})(m_{1}^{\prime}+m_{1}^{\prime\prime})\Bigg]\Bigg\}.\label{eq:fplusfminus}
\end{eqnarray}
Finally we get the form factors   through  the  
relations:
\begin{equation}
F_{1}^{PP}(q^{2})=f_{+}(q^{2}),\quad F_{0}^{PP}(q^{2})=f_{+}(q^{2})+\frac{q^{2}}{q\cdot P}f_{-}(q^{2}).
\end{equation}
Similarly, one can derive  the other  form factors,  whose expressions are collected  in Appendix \ref{app:FF}.

Before closing this section, it is worth mentioning that the axial-vector
mesons may not be   classified as $\,^{3}P_{1}$ or $\,^{1}P_{1}$
state. In the quark limit with $m_{Q}\to \infty$, the QCD interaction is independent of the heavy quark spin and thus it will   decouple with the light system. A  consequence of this decoupling is that heavy mesons are classified into  multiplets labeled by the total angular momentum of the light degrees of freedom. The s-wave  pseudo-scalar and vector states are in the same multiplets denoted as $s_l=1/2$. For the p-wave states,   two kinds of axial-vector mesons $P^{3/2}_1$ and   $P^{1/2}_1$ are mixtures of $\,^{3}P_{1}$ or $\,^{1}P_{1}$: 
\begin{eqnarray}
 |P^{3/2}_1\rangle &=& \sqrt{\frac{2}{3}} |^1P_1\rangle +\sqrt{\frac{1}{3}}
 |^3P_1\rangle, \;\;\;
 |P^{1/2}_1\rangle = \sqrt{\frac{1}{3}} |^1P_1\rangle -\sqrt{\frac{2}{3}}
 |^3P_1\rangle.
\end{eqnarray}
Since the form factors involving $P_{1}^{3/2}$ and
$P_{1}^{1/2}$ can be straightforwardly  obtained by the linear combination for those
given above,   we shall calculate the form factors using the $^{2S+1}L_J$ basis in the following analysis.

%%%%%%%%%%%%%%%%%%%%%%%%
\section{NUMERICAL RESULTS FOR   FORM FACTORS}
\label{sec:numerics}
%%%%%%%%%%%%%%%%%%%%%%%%

\subsection{Input parameters}

In the covariant LFQM,
the constituent quark masses are used as (in units of GeV):
\begin{equation}
m_{u}=m_{d}=0.25,\quad m_{s}=0.37,\quad m_{c}=1.4,\;\;\;m_{b}=4.8,
\end{equation}
which have been widely used in various $B$ and $B_c$ decays~\cite{Lu:2007sg,Wang:2007sxa,Wang:2008xt,Wang:2008ci,Wang:2009mi,Chen:2009qk,Li:2010bb,Verma:2011yw}.
The masses of the $B_c$ and $B_{sJ}$  are taken from the PDG  (in units of GeV)~\cite{Agashe:2014kda}:
\begin{eqnarray}
m_{B_{c}}=6.276,\;\;\; m_{B_{s}}=5.367,\;\;\; m_{B_{s}^{*}}=5.415, \;\;\; m_{B_{s2}}=5.840,\label{eq:massBcBs}
\end{eqnarray}
while for the  $B_{s0}$ and $B_{s1}$,  we quote  the  results~\cite{Liu:2015uya,Liu:2016efm} (see also estimates in Refs.~\cite{DiPierro:2001dwf,Ebert:2009ua,Wang:2012pf}):
\begin{eqnarray}
m_{B_{s0}}=5.782,\;\;\;   m_{B_{s1}(P_1^{1/2})}=5.843,\;\;\;   m_{B_{s1}(P_1^{3/2})}=5.833.\label{eq:massBs01}
\end{eqnarray}
Since masses of the $P_1^{1/2}$ and $P_1^{3/2}$ are close to the observed state  $B_{s1}(5830)$~\cite{Agashe:2014kda},
\begin{eqnarray}
 m_{B_{s1}(5830)}= 5.829 {\rm GeV},\label{eq:massBs1}
\end{eqnarray}
we use the same value for both the $^3P_1$ and $^1P_1$ state.

The   parameter $\beta$, characterizing the momentum distribution,
is usually determined  by fitting the meson decay constant. For instance, in this approach
the pseudoscalar and vector meson\textquoteright s decay constants
read
\begin{eqnarray}
f_{P} & = & \frac{N_{c}}{16\pi^{3}}\int dx_{2}d^{2}p_{\perp}^{\prime}\frac{h_{P}^{\prime}}{x_{1}x_{2}(M^{\prime2}-M_{0}^{\prime2})}4(m_{1}^{\prime}x_{2}+m_{2}x_{1}^{\prime}),\nonumber \\
f_{V} & = & \frac{N_{c}}{4\pi^{3}M^{\prime}}\int dx_{2}d^{2}p_{\perp}^{\prime}\frac{h_{V}^{\prime}}{x_{1}x_{2}(M^{\prime2}-M_{0}^{\prime2})} \left[x_{1}M_{0}^{\prime2}-m_{1}^{\prime}(m_{1}^{\prime}-m_{2})-p_{\perp}^{\prime2}+\frac{m_{1}^{\prime}+m_{2}}{w_{V}^{\prime}}p_{\perp}^{\prime2}\right].
\end{eqnarray}

For the $B_c$ meson, the decay constant can  be in principle determined by   leptonic and radiative-leptonic decays~\cite{Chang:1997re,Chang:1999gn,Chen:2015csa,Wang:2015bka}, both of which are  lack of experimental  data yet. Two loop contributions in the NRQCD framework have been calculated in Ref.~\cite{Chen:2015csa} and the authors have found:
\begin{eqnarray}
f_{B_{c}} & = & 398{\rm MeV}.
\end{eqnarray}
We will adopt this result, but it is necessary to note the above value is  smaller   than  Lattice QCD result by approximately  2$\sigma$:  $f_{B_{c}} = (434\pm15){\rm MeV}$.
We use the recent  Lattice QCD result for the $B_s$ decay constant with $N_f=2+1+1$~\cite{Bussone:2016iua}
\begin{eqnarray}
f_{B_{s}} & = & (229\pm 5){\rm MeV}.
\end{eqnarray}
This is close to the previous Lattice QCD result~\cite{Dowdall:2013tga,Aoki:2013ldr}: $f_{B_{s}} = (224\pm 5){\rm MeV}$.
Using the decay constants, the shape parameters  are fixed as  
\begin{eqnarray}
\beta_{B_{c}} & = & 0.886 {\rm GeV},  \;\; \beta_{B_{s}}  =  0.623 {\rm GeV},
\end{eqnarray}
and we assume that the values of $\beta$ for other $B_{sJ}$ mesons
are approximately equal to that  for the $B_{s}$, that is
\begin{equation}
\beta_{B_{s}^{*}}=\beta_{B_{s0}}=\beta_{B_{s1}}=\beta_{B_{s1}^{\prime}}=\beta_{B_{s2}}=0.623 {\rm GeV}.
\end{equation}
%To account for the inadequate knowledge on decay constants of the $B_{sJ}$, we shall  introduce $10\%$ errors to the $\beta$s.

We will also calculate the $B_c\to B_J$ form factors, for which we use the masses~\cite{Liu:2015uya,Liu:2016efm}
\begin{eqnarray}
 m_{B}= 5.279{\rm GeV},\;\;
 m_{B^*}= 5.325{\rm GeV}, \;\;
 m_{B_0}= 5.749{\rm GeV}, \;\;
 m_{B_1}= m_{B_1'}=  5.731 {\rm GeV}, \;\;
 m_{B_2}= 5.746{\rm GeV},\label{eq:massB}
\end{eqnarray}
and the shape parameter  $\beta$ for the $B_J$ meson:
\begin{eqnarray}
\beta_{B_J}= 0.562 {\rm GeV}. 
\end{eqnarray}
The above result is derived from  decay constant result~\cite{Bussone:2016iua}:
\begin{eqnarray}
f_{B}=   (193 \pm 6){\rm MeV}.
\end{eqnarray}

\subsection{Form factors and momentum transfer distribution}

%%%%%%%%%%%%%%%%%%%%%%%%%%%%%%%%%%%%%%%%%%%%%%%%%%%%%%%%%
\begin{table}
\caption{$B_{c}\to B_{s},B_{s}^{*},B_{s0},B_{s1},B_{s1}^{\prime}$ and
$B_{s2}$ form factors in the light-front quark model, which are fitted using equation ~\eqref{eq:fitminus} while for the form factors with an asterisk, the parametrization in Eq.~\eqref{eq:fitplus} is adopted.}
\label{Tab:BcBsformFactor}
\begin{center}
\begin{tabular}{c|c|c|c|c|c|c|c}
\hline\hline
$F$ &$F(0)$ &$m_{\rm{fit}}$ &$\delta$ &$F$ &$F(0)$ &$m_{\rm{fit}}$ &$\delta$\\
\hline
$F_{1}^{B_cB_s}$&$ 0.73$ &$ 1.57$ &$ 0.49$ &$F_{0}^{B_cB_s}$&$ 0.73$ &$ 2.07$ &$ 0.82$\\

$V^{B_cB_s^{*}}$&$ 3.70$ &$ 1.57$ &$ 0.48$ &$A_{0}^{B_cB_s^{*}}$&$ 0.55$ &$1.49$ &$ 0.61$\\

$A_{1}^{B_cB_s^{*}}$&$ 0.52$ &$ 1.90$ &$ 0.56$ &$A_{2}^{B_cB_s^{*}}$&$ 0.07^*$ &$ 1.04^*$ &$ 0.37^*$\\

$F_{1}^{B_cB_{s0}}$&$ 0.71$ &$ 1.69$ &$ 0.48$ &$F_{0}^{B_cB_{s0}}$&$ 0.72^*$ &$ 1.98^*$ &$ 1.43^*$\\

$A^{B_cB_{s1}}$&$ 0.19$ &$ 1.71$ &$ 0.45$ &$V_{0}^{B_cB_{s1}}$&$ 0.10^*$ &$ 0.75^*$ &$ 0.95^*$\\

$V_{1}^{B_cB_{s1}}$&$ 5.28^*$ &$ 2.28^*$ &$ 2.08^*$ &$V_{2}^{B_cB_{s1}}$&$ 0.07$ &$1.73$ &$ 0.32$\\

$A^{B_cB_{s1}^{\prime}}$&$ 0.05$ &$ 1.58$ &$ 0.51$ &$V_{0}^{B_cB_{s1}^{\prime}}$&$ 0.63$ &$ 1.76$ &$ 0.60$\\

$V_{1}^{B_cB_{s1}^{\prime}}$&$ 10.30$ &$ 1.71$ &$ 0.48$ &$V_{2}^{B_cB_{s1}^{\prime}}$&$-0.23$ &$ 1.49$ &$ 0.49$\\

$V^{B_cB_{s2}}$&$-18.60$ &$ 1.50$ &$ 0.48$ &$A_{0}^{B_cB_{s2}}$&$-2.94$ &$1.47$ &$ 0.54$\\

$A_{1}^{B_cB_{s2}}$&$-2.89$ &$ 1.75$ &$ 0.48$ &$A_{2}^{B_cB_{s2}}$&$-1.32^*$ &$3.24^*$ &$ 9.56^*$\\
\hline\hline
\end{tabular}
\end{center}
\end{table}
%%%%%%%%%%%%%%%%%%%%%%%%%%%%%%%%%%%%%%%%%%%%%%%%%%%%%%%%%

%%%%%%%%%%%%%%%%%%%%%%%%%%%%%%%%%%%%%%%%%%%%%%%%%%%%%%%%%
\begin{table}
\caption{$B_{c}\to B,B^{*},B_{0},B_{1},B_{1}^{\prime}$ and
$B_{2}$ form factors in the covariant LFQM  fitted through Eq.~\eqref{eq:fitminus},  except for the form factors with an asterisk, which are fitted using Eq.~\eqref{eq:fitplus}.  }
\label{Tab:BcBformFactor}
\begin{center}
\begin{tabular}{c|c|c|c|c|c|c|c}
\hline\hline
$F$ &$F(0)$ &$m_{\rm{fit}}$ &$\delta$ &$F$ &$F(0)$ &$m_{\rm{fit}}$ &$\delta$\\ \hline

$F_{1}^{B_cB}$&$ 0.64$ &$ 1.50$ &$ 0.52$ &$F_{0}^{B_cB}$&$ 0.64$ &$ 1.94$ &$0.83$\\

$V^{B_cB^{*}}$&$ 3.44$ &$ 1.50$ &$ 0.51$ &$A_{0}^{B_cB^{*}}$&$ 0.47$ &$ 1.42$ &$ 0.68$\\

$A_{1}^{B_cB^{*}}$&$ 0.44$ &$ 1.84$ &$ 0.63$ &$A_{2}^{B_cB^{*}}$&$ 0.07^*$ &$1.03^*$ &$ 0.37^*$\\

$F_{1}^{B_cB_{0}}$&$ 0.69$ &$ 1.61$ &$ 0.51$ &$F_{0}^{B_cB_{0}}$&$ 0.69^*$ &$2.83^*$ &$ 4.84^*$\\

$A^{B_cB_{1}}$&$ 0.21$ &$ 1.64$ &$ 0.49$ &$V_{0}^{B_cB_{1}}$&$ 0.13^*$ &$ 2.48^*$ &$ 51.50^*$\\

$V_{1}^{B_cB_{1}}$&$ 4.97^*$ &$ 3.14^*$ &$ 6.49^*$ &$V_{2}^{B_cB_{1}}$&$ 0.09$ &$1.64$ &$ 0.38$\\

$A^{B_cB_{1}^{\prime}}$&$ 0.06$ &$ 1.51$ &$ 0.55$ &$V_{0}^{B_cB_{1}^{\prime}}$&$ 0.64$ &$ 1.66$ &$ 0.64$\\

$V_{1}^{B_cB_{1}^{\prime}}$&$ 8.05$ &$ 1.62$ &$ 0.51$ &$V_{2}^{B_cB_{1}^{\prime}}$&$-0.24$ &$ 1.42$ &$ 0.53$\\

$V^{B_cB_{2}}$&$17.60$ &$ 1.43$ &$ 0.52$ &$A_{0}^{B_cB_{2}}$&$2.64$ &$ 1.40$ &$ 0.59$\\

$A_{1}^{B_cB_{2}}$&$2.59$ &$ 1.68$ &$ 0.52$ &$A_{2}^{B_cB_{2}}$&$1.31^*$ &$3.13^*$ &$ 9.72^*$\\
\hline\hline
\end{tabular}
\end{center}
\end{table}
%%%%%%%%%%%%%%%%%%%%%%%%%%%%%%%%%%%%%%%%%%%%%%%%%%%%%%%%%

With the inputs in the previous subsection, we can predict the   $B_{c}\to B_{s},B_{s}^{*},B_{s0},B_{s1},B_{s1}^{\prime}$
and $B_{s2}$ form factors in the LFQM and 
we show our results in Table~\ref{Tab:BcBsformFactor}. In order to access the $q^2$ distribution,  one may adopt  the fit formula:
\begin{equation}
F(q^{2})=\frac{F(0)}{1-\frac{q^{2}}{m_{\text{fit}}^{2}}+\delta(\frac{q^{2}}{m_{\text{fit}}^{2}})^{2}}.\label{eq:fitminus}
\end{equation}

In the literature, the dipole form has been used to parametrize the $q^2$ distribution: 
\begin{eqnarray}
 F(q^2)=\frac{F(0)}{1-a\frac{q^{2}}{m_{H}^{2}}+b(\frac{q^{2}}{m_{H}^{2}})^{2}},\label{eq:fit_Bpi}
\end{eqnarray}
with the $m_H=m_D$ for $D$ decays and $m_{H}=m_{B}$ for $B$ decays.  This parametrization is inspired by the analyticity. Taking the $F_1^{B\to \pi}$ as an example, we consider the timelike matrix element: 
\begin{eqnarray}
 \langle 0| \bar u\gamma^\mu b|\pi(-p_\pi) \overline B(p_{B})\rangle \sim \int \frac{d^4q}{(2\pi)^4}\frac{i}{q^2-m_X^2}  \langle 0|\bar u\gamma^\mu b|X\rangle \langle X| \pi(-p_\pi) \overline B(p_{B})\rangle,
\end{eqnarray}
where the one-particle contribution has been singled out. 
The lowest resonance that can contribute is the vector $B^*$. This leads  to the pole structure at   large $q^2$:  
\begin{eqnarray}
 F_1^{B\to \pi} (q^2) \sim \frac{F_1(0)}{1-q^2/m_{B^*}^2}. 
\end{eqnarray}
Except the pole  at $m_{B^*}$, there are residual dependences on $q^2$ which can be effectively incorporated into the   $a,b$ of the dipole parametrization as shown in Eq.~\eqref{eq:fit_Bpi}. 
However for the $B_c\to B_{sJ}$ transition, one can not simply apply  Eq.~\eqref{eq:fit_Bpi}, since the contributing  states are the $D_{s}$ resonances.  Using the $m_{H}=m_{B_c}$ will not only  disguise the genuine poles, but also lead to  irrationally  large results for   parameters  $a$ and $b$.   So in order to avoid this problem, we have adopted the parametrization in Eq.~\eqref{eq:fitminus}. From the results in Tab.~\ref{Tab:BcBsformFactor}, one can see that   the $m_{fit}$ for most form factors is between $1.5$ GeV to $2.0$ GeV, close to the mass of a    $D_{sJ}$ resonance.  This has validated  our parametrization.

For the $A_{2}^{B_cB_s^{*}}$, $F_{0}^{B_cB_{s0}}$, $V_{0,1}^{B_cB_{s1}}$ and $A_{2}^{B_cB_{s2}}$, we found that the fitted values for the $m_{fit}^2$ are negative, and thus we use  the following formula:
\begin{equation}
F(q^{2})=\frac{F(0)}{1+\frac{q^{2}}{m_{\text{fit}}^{2}}+\delta(\frac{q^{2}}{m_{\text{fit}}^{2}})^{2}}.\label{eq:fitplus}
\end{equation}

\begin{figure}
\begin{center}
\includegraphics[scale=0.8]{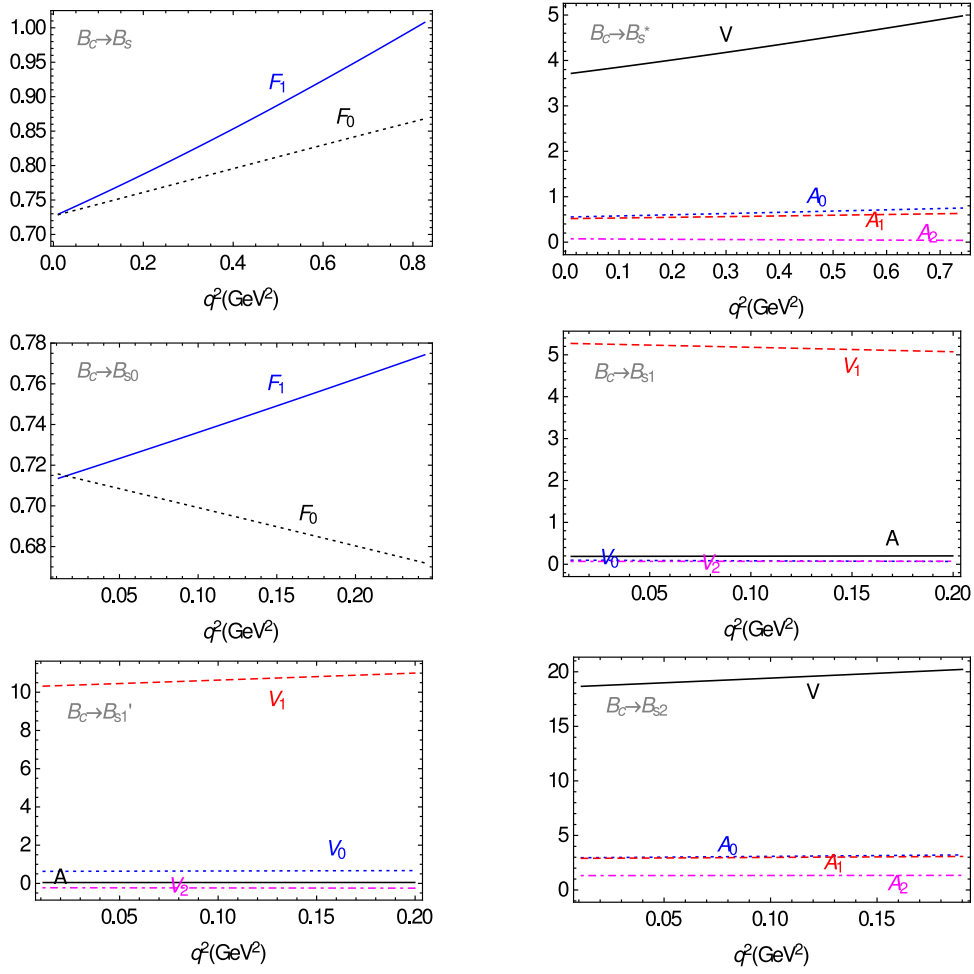}
\caption{The $B_c \to B_s$ form factors and their $q^2$-dependence parameterized by Eq.~\eqref{eq:fitminus} and Eq.~\eqref{eq:fitplus}.   } \label{fig:ffBcBs}
\end{center}
\end{figure}

The $q^2$-dependent form factors of $B_c \to B_s $  are shown in Fig.~\ref{fig:ffBcBs}.  From this figure, we can see that except for the $B_c\to B_{s1}$ transition, most form factors are rather stable against the variation of $q^2$.  This is partly because of the limited phase space. This will also lead to a reliable prediction for the branching fractions given in the next section.

%%%%%%%%%%%%%%%%%%%%%%%%%%%%%%%%%%%%%%%%%%%%%%%%%%%%
%%%%%%%%%%%%%%%%%%%%%%%%%%%%%%%%%%%%%%%%%%%%%%%%%%%%%

%\subsection{Comparison with other model calculations}

%It is useful to compare our results based on the covariant
%light-front model with other theoretical calculations.

%%%%%%%%%%%%%%%%%%%%%%%%%%%%%%%%%%%%%%%%%%%%%%%%%%%%%%%%%%%%%
%%%%%%%%%%%%%%%%%%%%%%%%%%%%%%%%%%%%%%%%%%%%%%%%%%%%%%%%%%%%%
\section{Phenomenological applications}
\label{sec:application}
%%%%%%%%%%%%%%%%%%%%%%%%%%%%%%%%%%%%%%%%%%%%%%%%%%%%%%%%%%%%%
%%%%%%%%%%%%%%%%%%%%%%%%%%%%%%%%%%%%%%%%%%%%%%%%%%%%%%%%%%%%%

\subsection{Semileptonic $B_{c}$ decays}

Decay width for semileptonic decays of $B_{c}\to M\bar{l}\nu$, where
$M=P,V,S,A,T$, can be derived by dividing the decay amplitude into
hadronic part and leptonic part, both of which are Lorentz invariant
so that can be readily evaluated. Then the differential decay widths
for $B_{c}\to P\bar{l}\nu$ and $B_{c}\to V\bar{l}\nu$ turn out
to be 
\begin{eqnarray}
\frac{d\Gamma(B_{c}\to P\bar{l}\nu)}{dq^{2}} & = & \left(1-\hat m_{l}^{2} \right)^{2}\frac{\sqrt{\lambda(m_{B_{c}}^{2},m_{P}^{2},q^{2})}G_{F}^{2}|V_{{\rm CKM}}|^{2}}{384m_{B_{c}}^{3}\pi^{3}}\Bigg\{(\hat m_{l}^{2}+2)\lambda(m_{B_{c}}^{2},m_{P}^{2},q^{2})F_{1}^{2}(q^{2})\nonumber \\
 &  & +3\hat m_{l}^{2}(m_{B_{c}}^{2}-m_{P}^{2})^{2}F_{0}^{2}(q^{2})\Bigg\},
 \end{eqnarray}
 %%%%%%%%%%%%
 \begin{eqnarray}
\frac{d\Gamma_{L}(B_{c}\to V\bar{l}\nu)}{dq^{2}} & = & \left(1-\hat m_{l}^{2}\right)^{2}\frac{\sqrt{\lambda(m_{B_{c}}^{2},m_{V}^{2},q^{2})}G_{F}^{2}|V_{{\rm CKM}}|^{2}}{384m_{B_{c}}^{3}\pi^{3}} \Bigg\{3\hat m_{l}^{2}\lambda(m_{B_{c}}^{2},m_{V}^{2},q^{2})A_{0}^{2}(q^{2})+(\hat m_{l}^{2}+2)\nonumber \\
 &  & \times\left|\frac{1}{2m_{V}}\left[(m_{B_{c}}^{2}-m_{V}^{2}-q^{2})(m_{B_{c}}+m_{V})A_{1}(q^{2})-\frac{\lambda(m_{B_{c}}^{2},m_{V}^{2},q^{2})}{m_{B_{c}}+m_{V}}A_{2}(q^{2})\right]\right|^{2}\Bigg\},\label{eq:GammaL}
 \end{eqnarray}
 %%%%%%%%%%%% %%%
 \begin{eqnarray}
\frac{d\Gamma^{\pm}(B_{c}\to V\bar{l}\nu)}{dq^{2}} & = & \left(1-\hat m_{l}^{2}\right)^{2}\frac{\sqrt{\lambda(m_{B_{c}}^{2},m_{V}^{2},q^{2})}G_{F}^{2}|V_{{\rm CKM}}|^{2}}{384m_{B_{c}}^{3}\pi^{3}}\Bigg\{(m_{l}^{2}+2q^{2})\lambda(m_{B_{c}}^{2},m_{V}^{2},q^{2})\nonumber \\
 &  & \times\left|\frac{V(q^{2})}{m_{B_{c}}+m_{V}}\mp\frac{(m_{B_{c}}+m_{V})A_{1}(q^{2})}{\sqrt{\lambda(m_{B_{c}}^{2},m_{V}^{2},q^{2})}}\right|^{2}\Bigg\},\label{eq:Gamma-plusminus}
\end{eqnarray}
where the superscript $+(-)$ denotes the right-handed (left-handed)
polarizations of vector mesons. $\lambda(m_{B_{c}}^{2},m_{i}^{2},q^{2})=(m_{B_{c}}^{2}+m_{i}^{2}-q^{2})^{2}-4m_{B_{c}}^{2}m_{i}^{2}$
with $i=P,V$. $\hat m_l= m_l/\sqrt{q^2}$.  The combined transverse and total differential decay
widths are given by
\begin{equation}
\frac{d\Gamma_{T}}{dq^{2}}=\frac{d\Gamma_{+}}{dq^{2}}+\frac{d\Gamma_{-}}{dq^{2}},\quad\quad\frac{d\Gamma}{dq^{2}}=\frac{d\Gamma_{L}}{dq^{2}}+\frac{d\Gamma_{T}}{dq^{2}}.
\end{equation}

The  differential decay widths for $B_{c}\to S\bar{l}\nu$
and $B_{c}\to A\bar{l}\nu$ can be obtained by making the following
replacements in the above expressions for $B_{c}\to P\bar{l}\nu$
and $B_{c}\to V\bar{l}\nu$
\begin{eqnarray}
m_{P} & \to & m_{S},\nonumber \\
F_{i}^{B_{c}P}(q^{2}) & \to & F_{i}^{B_{c}S}(q^{2}),\quad i=0,1
\end{eqnarray}
and
\begin{eqnarray}
m_{B_{c}}+m_{V} & \to & m_{B_{c}}-m_{A},\nonumber \\
V^{B_{c}V}(q^{2}) & \to & A^{B_{c}A}(q^{2}),\\
A_{i}^{B_{c}V}(q^{2}) & \to & V_{i}^{B_{c}A}(q^{2}),\quad i=0,1,2
\end{eqnarray}
respectively.
The  $d\Gamma_{L}/dq^{2}$ and $d\Gamma^{\pm}/dq^{2}$
for $B_{c}\to T\bar{l}\nu$ is  given by equation (\ref{eq:GammaL})
multiplied $(\sqrt{\frac{2}{3}}\frac{|\vec{p}_{T}|}{m_{T}})^{2}$
and Eq. (\ref{eq:Gamma-plusminus}) multiplied $(\frac{1}{\sqrt{2}}\frac{|\vec{p}_{T}|}{m_{T}})^{2}$,
respectively. Here the  $\vec{p}_{T}$ denotes the momentum of the tensor
meson in the $B_{c}$ rest frame and $m_{T}$ is mass of the tensor
meson.

For the  $B_{sJ}$ final state,  the inputs are   form
factors given in table \ref{Tab:BcBsformFactor} and the masses of $B_{c}$ and $B_{sJ}$s
given in Eqs. (\ref{eq:massBcBs}).  The other input parameters  are given as
follows \cite{Agashe:2014kda}:
\begin{eqnarray}
 &  & \tau_{B_{c}}=(0.452\times10^{-12}){\rm s},\nonumber \\
 &  & m_{e}=0.511{\rm MeV},\quad m_{\mu}=0.106{\rm GeV},\quad m_{\tau}=1.78{\rm GeV},\nonumber \\
 &  & G_{F}=1.166\times10^{-5}{\rm GeV}^{-2},\quad|V_{cs}|=0.973,\label{eq:input-semi}
\end{eqnarray}
and our predictions for branching fractions are given in Table \ref{Tab:BR-semiBcBsJ}.   It should be mentioned that in the above calculation we have considered
$B_{s1}(B_{s1}^{\prime})$ and $B_{1}(B_{1}^{\prime})$ to be in $\,^{3}P_{1}(\,^{1}P_{1})$
eigenstates.

For the $B_{J}$ final state, we need to evaulate  the form factors for $B_c \to B_{J}$ by following the same method, and our results are given in Table \ref{Tab:BcBformFactor}.  The masses of $B_{c}$ and $B_{J}$s are also 
given in Eqs. (\ref{eq:massBcBs}) and Eqs. (\ref{eq:massB}). The other inputs   are the same
as Eq.~\eqref{eq:input-semi} but with $|V_{cs}|=0.973$ replaced
by $|V_{cd}|=0.225$ \cite{Agashe:2014kda}. With these inputs,  our results for branching fractions and ratios are given in Table \ref{Tab:BR-semiBcBJ}.

From these tables, we can see that the branching fractions for $B_c\to B_s\bar \ell\nu$ and $B_c\to B_s^* \bar \ell\nu$ are at the percent level, while those for the $B_c\to  B\bar\ell\nu$ and $B_c\to  B^* \bar\ell\nu$ are suppressed by one order of magnitude.  This is consistent with the results in the literature~\cite{Chang:1992pt,Lu:1995ug,Liu:1997hr,Colangelo:1999zn,Ivanov:2000aj,Kiselev:2000pp,AbdElHady:1999xh,Choudhury:2001ty,Ebert:2003wc}.    Branching fractions for channels with $p$-wave bottomed mesons in the final state range from  $10^{-4}$ to $10^{-6}$. In decays with large phase space, the electron and muon masses can introduce about a few percents to branching ratios. While for those limited phase space like the $B_c\to B_{s2}\bar \ell\nu$, the effects due to the lepton mass difference can reach $30\%$.  We hope these predictions can be examined in future on the experimental side. 

%%%%%%%%%%%%%%%%%%%%%%%%%%%%%
\begin{table}
\caption{Branching fractions  for  $B_{c} \to B_{sJ} \bar{\ell} \nu$ using  the $B_{c} \to B_{sJ}$ form factors given in Table~\ref{Tab:BcBsformFactor}. Here $\ell=e,\mu$.    }
\label{Tab:BR-semiBcBsJ}
\begin{center}
\begin{tabular}{|c|c|c||c|c|c|}
\hline 
$\ell=e$ & ${\cal B}_{{\rm total}}$ & ${\cal B}_{L}/{\cal B}_{T}$& $\ell=\mu$ & ${\cal B}_{{\rm total}}$ & ${\cal B}_{L}/{\cal B}_{T}$\tabularnewline
\hline 
\hline 
$B_{c}\to B_{s}\bar{\ell}\nu$ & $1.51\times10^{-2}$ & $--$ & $B_{c}\to B_{s}\bar{\ell}\nu$ & $1.43\times10^{-2}$ & $--$\tabularnewline
\hline 
$B_{c}\to B_{s}^{*}\bar{\ell}\nu$ & $1.96\times10^{-2}$ & $1.13$&$B_{c}\to B_{s}^{*}\bar{\ell}\nu$ & $1.83\times10^{-2}$ & $1.10$ \tabularnewline
\hline 
$B_{c}\to B_{s0}\bar{\ell}\nu$ & $6.58\times10^{-4}$ & $--$&$B_{c}\to B_{s0}\bar{\ell}\nu$ & $5.23\times10^{-4}$ & $--$\tabularnewline
\hline 
$B_{c}\to B_{s1}\bar{\ell}\nu$ & $8.31\times10^{-5}$ & $0.57$&
$B_{c}\to B_{s1}\bar{\ell}\nu$ & $6.33\times10^{-5}$ & $0.52$\tabularnewline
\hline 
$B_{c}\to B_{s1}^{\prime}\bar{\ell}\nu$ & $5.38\times10^{-4}$ & $2.38$&
$B_{c}\to B_{s1}^{\prime}\bar{\ell}\nu$ & $3.98\times10^{-4}$ & $2.09$\tabularnewline
\hline 
$B_{c}\to B_{s2}\bar{\ell}\nu$ & $2.98\times10^{-5}$ & $2.29$&
$B_{c}\to B_{s2}\bar{\ell}\nu$ & $1.97\times10^{-5}$ & $1.97$\tabularnewline
\hline 
\end{tabular}
\end{center}
\end{table}
%%%%%%%%%%%%%%%%%%%%%%%%%%%%%

%%%%%%%%%%%%%%%%%%%%%%%%%%%%%
\begin{table}
\caption{Branching ratios for $B_{c} \to B_{J} \bar{\ell} \nu (\ell=e,\mu)$ with the $B_{c} \to B_{J}$ form factors given in Table \ref{Tab:BcBformFactor}.  }
\label{Tab:BR-semiBcBJ}
\begin{center}
\begin{tabular}{|c|c|c||c|c|c|}
\hline 
$\ell=e$ & ${\cal B}_{{\rm total}}$ & ${\cal B}_{L}/{\cal B}_{T}$& $\ell=\mu$ & ${\cal B}_{{\rm total}}$ & ${\cal B}_{L}/{\cal B}_{T}$\tabularnewline
\hline 
\hline 
$B_{c}\to B\bar{\ell}\nu$ & $1.04\times10^{-3}$ & $--$ & $B_{c}\to B\bar{\ell}\nu$ & $1.00\times10^{-3}$ & $--$\tabularnewline
\hline 
$B_{c}\to B^{*}\bar{\ell}\nu$ & $1.34\times10^{-3}$ & $1.06$&$B_{c}\to B^{*}\bar{\ell}\nu$ & $1.27\times10^{-3}$ & $1.04$ \tabularnewline
\hline 
$B_{c}\to B_{0}\bar{\ell}\nu$ & $4.60\times10^{-5}$ & $--$&$B_{c}\to B_{0}\bar{\ell}\nu$ & $3.77\times10^{-5}$ & $--$\tabularnewline
\hline 
$B_{c}\to B_{1}\bar{\ell}\nu$ & $1.52\times10^{-5}$ & $0.52$&
$B_{c}\to B_{1}\bar{\ell}\nu$ & $1.28\times10^{-5}$ & $0.50$\tabularnewline
\hline 
$B_{c}\to B_{1}^{\prime}\bar{\ell}\nu$ & $7.70\times10^{-5}$ & $2.51$&
$B_{c}\to B_{1}^{\prime}\bar{\ell}\nu$ & $6.28\times10^{-5}$ & $2.29$\tabularnewline
\hline 
$B_{c}\to B_{2}\bar{\ell}\nu$ & $5.15\times10^{-6}$ & $2.22$&
$B_{c}\to B_{2}\bar{\ell}\nu$ & $3.90\times10^{-6}$ & $2.00$\tabularnewline
\hline 
\end{tabular}
\end{center}
\end{table}
%%%%%%%%%%%%%%%%%%%%%%%%%%%%%

%%%%%%%%%%%%%%%%%%%%%%%%%%%
\subsection{Nonleptonic $B_c$ decays}

Since our main purpose  of this work is to investigate the production
of $B_{sJ}$, we will focus on the decay modes which can be controlled
under the factorization approach. Such  decay modes are usually
dominated by tree operators with effective Hamiltonian
\begin{eqnarray}
{\cal H}_{{\rm eff}}(c\to su\bar{d}) & = & \frac{G_{F}}{\sqrt{2}}V_{cs}^{*}V_{ud}\Bigg\{ C_{1}[\bar{s}_{\alpha}\gamma^{\mu}(1-\gamma_{5})c_{\beta}][\bar{u}_{\beta}\gamma_{\mu}(1-\gamma_{5})d_{\alpha}]\nonumber \\
 &  & +C_{2}[\bar{s}_{\alpha}\gamma^{\mu}(1-\gamma_{5})c_{\alpha}][\bar{u}_{\beta}\gamma_{\mu}(1-\gamma_{5})d_{\beta}]\Bigg\},
\end{eqnarray}
where $C_{1}$ and $C_{2}$ are the Wilson coefficients, $\alpha$
and $\beta$ denote the color indices.

With the definitions of decay constants,
\begin{equation}
\langle\pi^{+}(p)|\bar{u}\gamma_{\mu}\gamma_{5}d|0\rangle=-if_{\pi}p_{\mu},
\end{equation}
one can expect the factorizaton formula to have the following forms
\begin{eqnarray}
i{\cal M}(B_{c}^{+}\to B_{s}\pi^{+}) & = & Nm_{B_{c}}^{2}(1-r_{B_{s}}^{2})F_{0}^{B_{c}B_{s}}(m_{\pi}^{2}),\\
i{\cal M}(B_{c}^{+}\to B_{s}^{*}\pi^{+}) & = & (-i)N\sqrt{\lambda(m_{B_{c}}^{2},m_{B_{s}^{*}}^{2},m_{\pi}^{2})}A_{0}^{B_{c}B_{s}^{*}}(m_{\pi}^{2}),\\
i{\cal M}(B_{c}^{+}\to B_{s0}\pi^{+}) & = & (-i)Nm_{B_{c}}^{2}(1-r_{B_{s0}}^{2})F_{0}^{B_{c}B_{s0}}(m_{\pi}^{2}),\\
i{\cal M}(B_{c}^{+}\to B_{s1}\pi^{+}) & = & (-i)N\sqrt{\lambda(m_{B_{c}}^{2},m_{B_{s1}}^{2},m_{\pi}^{2})}V_{0}^{B_{c}B_{s1}}(m_{\pi}^{2}),\\
i{\cal M}(B_{c}^{+}\to B_{s1}^{\prime}\pi^{+}) & = & (-i)N\sqrt{\lambda(m_{B_{c}}^{2},m_{B_{s1}^{\prime}}^{2},m_{\pi}^{2})}V_{0}^{B_{c}B_{s1}^{\prime}}(m_{\pi}^{2}),\\
i{\cal M}(B_{c}^{+}\to B_{s2}\pi^{+}) & = & (-i)\frac{1}{\sqrt{6}}N\frac{\lambda(m_{B_{c}}^{2},m_{B_{s2}}^{2},m_{\pi}^{2})}{m_{B_{c}}^{2}r_{B_{s2}}}A_{0}^{B_{c}B_{s2}}(m_{\pi}^{2}),
\end{eqnarray}
where $N= {G_{F}}/{\sqrt{2}}V_{cs}^{*}V_{ud}a_{1}f_{\pi}$, with
$a_{1}=C_{2}+ {C_{1}}/{N_{c}}(N_{c}=3)$.

The partical decay width for $B_{c}\to B_{sJ}\pi$ is given as
\begin{equation}
\Gamma=\frac{|\vec{p}_{1}|}{8\pi m_{B_{c}}^{2}}|{\cal M}|^{2}
\end{equation}
with $|\vec{p}_{1}|$ being the magnitude of three-momentum of $B_{sJ}$
or $\pi$ meson in the final state in the $B_{c}$ rest frame.

We use the transition form factors given in Table \ref{Tab:BcBsformFactor} and the masses
of $B_{c}$ and $B_{sJ}$s given in Eqs. (\ref{eq:massBcBs}), (\ref{eq:massBs01}) and (\ref{eq:massBs1})  and the other
inputs which are given as follows \cite{Agashe:2014kda,Wang:2008xt}:
\begin{eqnarray}
 &  & \tau_{B_{c}}=(0.452\times10^{-12}){\rm s},\quad m_{\pi}=0.140{\rm GeV},\nonumber \\
 &  & |V_{cs}|=0.973,\quad|V_{ud}|=0.974,\\
 &  & f_{\pi}=130.4{\rm MeV},\quad a_{1}=1.07,
\end{eqnarray}
where $f_{\pi}$ can be extracted from $\pi^{-}\to\ell^{-}\bar{\nu}$
data and $a_{1}$ is evaluated at the typical fatorization
scale $\mu\sim m_c$ \cite{Buchalla:1995vs}.Then
our theoretical  results for $B_{c}\to B_{sJ}\pi$ branching ratios turn
out to be as follows:
\begin{eqnarray}
{\cal B}(B_{c}^{+}\to B_{s}\pi^{+}) & = & 4.1\%,\nonumber\\
{\cal B}(B_{c}^{+}\to B_{s}^{*}\pi^{+}) & = & 2.0\%,\nonumber\\
{\cal B}(B_{c}^{+}\to B_{s0}\pi^{+}) & = & 0.68\%,\nonumber\\
{\cal B}(B_{c}^{+}\to B_{s1}\pi^{+}) & = & 0.0082\%,\nonumber\\
{\cal B}(B_{c}^{+}\to B_{s1}^{\prime}\pi^{+}) & = & 0.36\%,\nonumber\\
{\cal B}(B_{c}^{+}\to B_{s2}\pi^{+}) & = & 0.023\%.
\end{eqnarray}

Using the $1fb^{-1}$ data of proton-proton collisions  collected at the center-of-mass energy of 7 TeV and $2fb^{-1}$ data accumulated at 8 TeV,  the LHCb collaboration has observed the decay $B_c\to B_s\pi^+$~\cite{Aaij:2013cda}:
\begin{eqnarray}
\frac{\sigma(B_c^+)}{\sigma(B_s^0)} \times {\cal B}(B_c^+\to B_s^0\pi^+) = (2.37\pm0.31\pm0.11^{+0.17}_{-0.13})\times 10^{-3}. 
\end{eqnarray}
The first uncertainty is statistical, the second is systematic and the third
arises from the uncertainty on the $B_c^+$ lifetime.  The ratio of cross sections $ {\sigma(B_c^+)}/{\sigma(B_s^0)} $ depends significantly on the kinematics, and a rough  estimate has lead to the branching ratio for $B_c^+\to B_s^0\pi^+$ of about $10\%$~\cite{Aaij:2013cda}.  The estimated branching fraction is somewhat larger than  but still at the same magnitude with our result.  Moreover, our results have indicated that the LHCb collaboration might be able to discover other channels with similar branching fractions like the $B_c\to B_s^*\pi$.

%%%%%%%%%%%%%%%%%%%%%%%%%%%%
\section{Conclusions}
%%%%%%%%%%%%%%%%%%%%%%%%%%%%

%The study on the spectrum of the heavy-light
%mesons is not only valuable to understand the nonperturbative
%dynamics in the charm sector, but is also of prime importance to
%explore the exotic candidates beyond the naive quark model.

To understand the structure of the heavy-light mesons, especially
the newly observed states, and to establish an overview of the
spectroscopy, a lot of effort are requested  on both experiment and
theory sides. One particular remark  is the classification of these
states. In the heavy quark limit, the charm quark will decouple with
the light degree of freedom and acts as a static color source.
Strong interactions will be independent of the heavy flavor and
spin. In this case, heavy mesons, the eigenstates of the QCD
Lagrangian in the heavy quark limit, can be labeled according to the
total angular momentum $s_l$ of the light degree of freedom. The
heavy mesons with the same angular momentum $s_l$ but different
orientations of the heavy quark spin degenerate.   One consequence
is that heavy mesons can be classified by the multiplets characterized
by $s_l$ instead of the usual scheme using the $^{2S+1}L_J$.

In this work, we have suggested  to study the $B_{s}$ and its excitations   $B_{sJ}$ in the $B_c$ decays.
We have  calculated the $B_c\to B_{sJ}$ and $B_c\to B_{J}$ form factors within the covariant light-front quark model, where the $B_{sJ}$ and $B_{J}$ denotes an s-wave or p-wave $\bar bs$ and $\bar bd$ meson, respectively.  The form factors  at $q^2=0$ are directly calculated  while the $q^2$-distribution is obtained by the extrapolation.  The derived form factors are then used to study semileptonic $B_c\to (B_{sJ},B_{J})\bar\ell\nu$  decays,  and  nonleptonic $B_c\to B_{sJ}\pi$. Branching fractions and polarizations are  predicted, through which  we find that the predicted  branching fractions are sizable, especially at the LHC experiment and future high-energy $e^+e^-$ colliders with a high luminosity at the $Z$-pole.  The future experimental measurements are helpful to study the nonperturbative QCD dynamics  in the presence of a heavy spectator and also of great value for the spectroscopy study.

%%%%%%%%%%%%%%%%%%%%%%%%%%%%
\section*{ACKNOWLEDGEMENTS}
%%%%%%%%%%%%%%%%%%%%%%%%%%%%

This work is supported in part  by National  Natural  Science Foundation of China under Grant  No.11575110,  Natural  Science Foundation of Shanghai under Grant  No. 15DZ2272100 and No. 15ZR1423100,  by the Open Project Program of State Key Laboratory of Theoretical Physics, Institute of Theoretical Physics, Chinese  Academy of Sciences, China (No.Y5KF111CJ1), and  by   Scientific Research Foundation for   Returned Overseas Chinese Scholars, State Education Ministry.

%\begin{appendices}
\appendix
\section{EXPLICIT EXPRESSIONS FOR FORM FACTORS}
\label{app:FF}

In the LFQM, it is more convenient to adopt a new set of parametrization of form factors, with the relations:
\begin{eqnarray}
V^{B_{c}V}(q^{2}) & = & -(m_{B_{c}}+m_{V})g(q^{2}),\quad A_{1}^{B_{c}V}(q^{2})=-\frac{f(q^{2})}{m_{B_{c}}+m_{V}}, A_{2}^{B_{c}V}(q^{2}) =  (m_{B_{c}}+m_{V})a_{+}(q^{2}),\nonumber \\
A_{0}^{B_{c}V}(q^{2}) & = & \frac{m_{B_{c}}+m_{V}}{2m_{V}}A_{1}(q^{2})-\frac{m_{B_{c}}-m_{V}}{2m_{V}}A_{2}(q^{2})-\frac{ q^{2}}{2m_{V}}a_{-}(q^{2}),\\
F_{1}^{B_{c}S}(q^{2}) & = & -u_{+}(q^{2}),\quad F_{0}^{B_{c}S}(q^{2})=-u_{+}(q^{2})-\frac{q^{2}}{q\cdot P}u_{-}(q^{2}), \nonumber \\
A^{B_{c}A}(q^{2}) & = & -(m_{B_{c}}-m_{A})q(q^{2}),\quad V_{1}^{B_{c}A}(q^{2})=-\frac{\ell(q^{2})}{m_{B_{c}}-m_{A}},
V_{2}^{B_{c}A}(q^{2})  =  (m_{B_{c}}-m_{A})c_{+}(q^{2}),\nonumber \\ 
V_{0}^{B_{c}A}(q^{2}) & = & \frac{m_{B_{c}}-m_{A}}{2m_{T}}V_{1}(q^{2})-\frac{m_{B_{c}}+m_{A}}{2m_{A}}V_{2}(q^{2})-\frac{ q^{2}}{2m_{A}}c_{-}(q^{2}),\\ 
V^{B_{c}T}(q^{2}) & = & -m_{B_{c}}(m_{B_{c}}+m_{T})h(q^{2}), A_{1}^{B_{c}T}(q^{2})  =   -\frac{m_{B_{c}}}{m_{B_{c}}+m_{T}}k(q^{2}),
A_{2}^{B_{c}T}(q^{2})   =  m_{B_{c}}(m_{B_{c}}+m_{T})b_{+}(q^{2}),\nonumber \\
A_{0}^{B_{c}T}(q^{2}) & = & \frac{m_{B_{c}}+m_{T}}{2m_{T}}A_{1}(q^{2})-\frac{m_{B_{c}}-m_{T}}{2m_{T}}A_{2}(q^{2})-\frac{m_{B_{c}}q^{2}}{2m_{T}}b_{-}(q^{2}).
\end{eqnarray}

The analytic  expressions for $P\to P$ transition form factors in the covariant LFQM have been
given in Eq. (\ref{eq:fplusfminus}), while   for the $P\to V$
transition, they are given as follows \cite{Jaus:1999zv,Cheng:2003sm}:
\begin{eqnarray}
g(q^{2}) & = & -\frac{N_{c}}{16\pi^{3}}\int dx_{2}d^{2}p_{\perp}^{\prime}\frac{2h_{P}^{\prime}h_{V}^{\prime\prime}}{x_{2}\hat{N}_{1}^{\prime}\hat{N}_{1}^{\prime\prime}}\Bigg\{ x_{2}m_{1}^{\prime}+x_{1}m_{2}+(m_{1}^{\prime}-m_{1}^{\prime\prime})\frac{p_{\perp}^{\prime}\cdot q_{\perp}}{q^{2}}
 +\frac{2}{w_{V}^{\prime\prime}}\Bigg[p_{\perp}^{\prime2}+\frac{(p_{\perp}^{\prime}\cdot q_{\perp})^{2}}{q^{2}}\Bigg]\Bigg\},
 \end{eqnarray}
 \begin{eqnarray}
f(q^{2}) & = & \frac{N_{c}}{16\pi^{3}}\int dx_{2}d^{2}p_{\perp}^{\prime}\frac{h_{P}^{\prime}h_{V}^{\prime\prime}}{x_{2}\hat{N}_{1}^{\prime}\hat{N}_{1}^{\prime\prime}}\Bigg\{2x_{1}(m_{2}-m_{1}^{\prime})(M_{0}^{\prime2}+M_{0}^{\prime\prime2})-4x_{1}m_{1}^{\prime\prime}M_{0}^{\prime2}\nonumber \\
 &  & +2x_{2}m_{1}^{\prime}q\cdot P+2m_{2}q^{2}-2x_{1}m_{2}(M^{\prime2}+M^{\prime\prime2})+2(m_{1}^{\prime}-m_{2})(m_{1}^{\prime}+m_{1}^{\prime\prime})^{2}\nonumber \\
 &  & +8(m_{1}^{\prime}-m_{2})\Bigg[p_{\perp}^{\prime2}+\frac{(p_{\perp}^{\prime}\cdot q_{\perp})^{2}}{q^{2}}\Bigg]+2(m_{1}^{\prime}+m_{1}^{\prime\prime})(q^{2}+q\cdot P)\frac{p_{\perp}^{\prime}\cdot q_{\perp}}{q^{2}}\nonumber \\
 &  & -4\frac{q^{2}p_{\perp}^{\prime2}+(p_{\perp}^{\prime}\cdot q_{\perp})^{2}}{q^{2}w_{V}^{\prime\prime}}\Bigg[2x_{1}(M^{\prime2}+M_{0}^{\prime2})-q^{2}-q\cdot P-2(q^{2}+q\cdot P)\frac{p_{\perp}^{\prime}\cdot q_{\perp}}{q^{2}}\nonumber \\
 &  & -2(m_{1}^{\prime}-m_{1}^{\prime\prime})(m_{1}^{\prime}-m_{2})\Bigg]\Bigg\},
 \end{eqnarray}
 \begin{eqnarray}
a_{+}(q^{2}) & = & \frac{N_{c}}{16\pi^{3}}\int dx_{2}d^{2}p_{\perp}^{\prime}\frac{2h_{P}^{\prime}h_{V}^{\prime\prime}}{x_{2}\hat{N}_{1}^{\prime}\hat{N}_{1}^{\prime\prime}}\Bigg\{(x_{1}-x_{2})(x_{2}m_{1}^{\prime}+x_{1}m_{2})
 -[2x_{1}m_{2}+m_{1}^{\prime\prime}+(x_{2}-x_{1})m_{1}^{\prime}]\frac{p_{\perp}^{\prime}\cdot q_{\perp}}{q^{2}}\nonumber \\
 &  & -2\frac{x_{2}q^{2}+p_{\perp}^{\prime}\cdot q_{\perp}}{x_{2}q^{2}w_{V}^{\prime\prime}}[p_{\perp}^{\prime}\cdot p_{\perp}^{\prime\prime}+(x_{1}m_{2}+x_{2}m_{1}^{\prime})(x_{1}m_{2}-x_{2}m_{1}^{\prime\prime})]\Bigg\},\end{eqnarray}
 \begin{eqnarray}
a_{-}(q^{2}) & = & \frac{N_{c}}{16\pi^{3}}\int dx_{2}d^{2}p_{\perp}^{\prime}\frac{h_{P}^{\prime}h_{V}^{\prime\prime}}{x_{2}\hat{N}_{1}^{\prime}\hat{N}_{1}^{\prime\prime}}\Bigg\{2(2x_{1}-3)(x_{2}m_{1}^{\prime}+x_{1}m_{2}) -8(m_{1}^{\prime}-m_{2})\Bigg[\frac{p_{\perp}^{\prime2}}{q^{2}}+2\frac{(p_{\perp}^{\prime}\cdot q_{\perp})^{2}}{q^{4}}\Bigg]\nonumber \\
 &  & -[(14-12x_{1})m_{1}^{\prime}-2m_{1}^{\prime\prime}-(8-12x_{1})m_{2}]\frac{p_{\perp}^{\prime}\cdot q_{\perp}}{q^{2}}\nonumber \\
 &  & +\frac{4}{w_{V}^{\prime\prime}}\Bigg([M^{\prime2}+M^{\prime\prime2}-q^{2}+2(m_{1}^{\prime}-m_{1}^{\prime\prime})(m_{1}^{\prime}-m_{2})](A_{3}^{(2)}+A_{4}^{(2)}-A_{2}^{(1)}) +Z_{2}(3A_{2}^{(1)}-2A_{4}^{(2)}-1)\nonumber \\
 &  & +\frac{1}{2}[x_{1}(q^{2}+q\cdot P)-2M^{\prime2}-2p_{\perp}^{\prime}\cdot q_{\perp}-2m_{1}^{\prime}(m_{1}^{\prime\prime}+m_{2}) -2m_{2}(m_{1}^{\prime}-m_{2})](A_{1}^{(1)}+A_{2}^{(1)}-1)\nonumber \\
 &  & +q\cdot P\Bigg[\frac{p_{\perp}^{\prime2}}{q^{2}}+\frac{(p_{\perp}^{\prime}\cdot q_{\perp})^{2}}{q^{4}}\Bigg](4A_{2}^{(1)}-3)\Bigg)\Bigg\}.
\end{eqnarray}

The explicit expressions for $P\to S$ and $P\to A$ transitions can be readily obtained by making the following replacements \cite{Cheng:2003sm}:
\begin{eqnarray}
u_{\pm}(q^{2}) & = & -f_{\pm}(q^{2})|_{m_{1}^{\prime\prime}\to-m_{1}^{\prime\prime},\; h_{P}^{\prime\prime}\to h_{S}^{\prime\prime}}, \nonumber\\
 {[} \ell^{\,^{3}A,\,^{1}A} (q^{2}) , q^{\,^{3}A,\,^{1}A}(q^{2}) ,  c_{\pm}^{\,^{3}A,\,^{1}A}(q^{2})  {]}& = & [f(q^{2}), g(q^2), a_{\pm}(q^2)]|_{ m_{1}^{\prime\prime}\to-m_{1}^{\prime\prime},\; h_{V}^{\prime\prime}\to h_{\,^{3}A,\,^{1}A}^{\prime\prime},\; w_{V}^{\prime\prime}\to w_{\,^{3}A,\,^{1}A}^{\prime\prime}},
\end{eqnarray}
where only the $1/W^{\prime\prime}$ terms in $P\to\,^{1}A$ form
factors are kept. It should be cautious that the replacement of $m_{1}^{\prime\prime}\to-m_{1}^{\prime\prime}$
should not be applied to $m_{1}^{\prime\prime}$ in $w^{\prime\prime}$
and $h^{\prime\prime}$.
The  $P\to T$ transition form factors are
calculated~\cite{Cheng:2003sm}
\begin{eqnarray}
h(q^{2}) & = & -g(q^{2})|_{h_{V}^{\prime\prime}\to h_{T}^{\prime\prime}}+\frac{N_{c}}{16\pi^{3}}\int dx_{2}d^{2}p_{\perp}^{\prime}\frac{2h_{P}^{\prime}h_{T}^{\prime\prime}}{x_{2}\hat{N}_{1}^{\prime}\hat{N}_{1}^{\prime\prime}}\Bigg[(m_{1}^{\prime}-m_{1}^{\prime\prime})(A_{3}^{(2)}+A_{4}^{(2)})\nonumber \\
 &  & +(m_{1}^{\prime\prime}+m_{1}^{\prime}-2m_{2})(A_{2}^{(2)}+A_{3}^{(2)})-m_{1}^{\prime}(A_{1}^{(1)}+A_{2}^{(1)})+\frac{2}{w_{V}^{\prime\prime}}(2A_{1}^{(3)}+2A_{2}^{(3)}-A_{1}^{(2)})\Bigg],\end{eqnarray}
\begin{eqnarray}
k(q^{2}) & = & -f(q^{2})|_{h_{V}^{\prime\prime}\to h_{T}^{\prime\prime}}+\frac{N_{c}}{16\pi^{3}}\int dx_{2}d^{2}p_{\perp}^{\prime}\frac{h_{P}^{\prime}h_{T}^{\prime\prime}}{x_{2}\hat{N}_{1}^{\prime}\hat{N}_{1}^{\prime\prime}}\Bigg\{2(A_{1}^{(1)}+A_{2}^{(1)})\nonumber \\
 &  & \times[m_{2}(q^{2}-\hat{N}_{1}^{\prime}-\hat{N}_{1}^{\prime\prime}-m_{1}^{\prime2}-m_{1}^{\prime\prime2})-m_{1}^{\prime}(M^{\prime\prime2}-\hat{N}_{1}^{\prime\prime}-m_{1}^{\prime\prime2}-m_{2}^{2})\nonumber \\
 &  & -m_{1}^{\prime\prime}(M^{\prime2}-\hat{N}_{1}^{\prime}-m_{1}^{\prime2}-m_{2}^{2})-2m_{1}^{\prime}m_{1}^{\prime\prime}m_{2}]+2(m_{1}^{\prime}+m_{1}^{\prime\prime})\Bigg(A_{2}^{(1)}Z_{2}+\frac{q\cdot P}{q^{2}}A_{1}^{(2)}\Bigg)\nonumber \\
 &  & +16(m_{2}-m_{1}^{\prime})(A_{1}^{(3)}+A_{2}^{(3)})+4(2m_{1}^{\prime}-m_{1}^{\prime\prime}-m_{2})A_{1}^{(2)}\nonumber \\
 &  & +\frac{4}{w_{V}^{\prime\prime}}\Bigg([M^{\prime2}+M^{\prime\prime2}-q^{2}+2(m_{1}^{\prime}-m_{2})(m_{1}^{\prime\prime}+m_{2})](2A_{1}^{(3)}+2A_{2}^{(3)}-A_{1}^{(2)})\nonumber \\
 &  & -4\Bigg[A_{2}^{(3)}Z_{2}+\frac{q\cdot P}{3q^{2}}(A_{1}^{(2)})^{2}\Bigg]+2A_{1}^{(2)}Z_{2}\Bigg)\Bigg\},\end{eqnarray}
\begin{eqnarray}
b_{+}(q^{2}) & = & -a_{+}(q^{2})|_{h_{V}^{\prime\prime}\to h_{T}^{\prime\prime}}+\frac{N_{c}}{16\pi^{3}}\int dx_{2}d^{2}p_{\perp}^{\prime}\frac{h_{P}^{\prime}h_{T}^{\prime\prime}}{x_{2}\hat{N}_{1}^{\prime}\hat{N}_{1}^{\prime\prime}}\Bigg\{8(m_{2}-m_{1}^{\prime})(A_{3}^{(3)}+2A_{4}^{(3)}+A_{5}^{(3)})\nonumber \\
 &  & -2m_{1}^{\prime}(A_{1}^{(1)}+A_{2}^{(1)})(A_{2}^{(2)}+A_{3}^{(2)})+2(m_{1}^{\prime}+m_{1}^{\prime\prime})(A_{2}^{(2)}+2A_{3}^{(2)}+A_{4}^{(2)})\nonumber \\
 &  & +\frac{2}{w_{V}^{\prime\prime}}[2[M^{\prime2}+M^{\prime\prime2}-q^{2}+2(m_{1}^{\prime}-m_{2})(m_{1}^{\prime\prime}+m_{2})](A_{3}^{(3)}+2A_{4}^{(3)}+A_{5}^{(3)}-A_{2}^{(2)}-A_{3}^{(2)})\nonumber \\
 &  & +[q^{2}-\hat{N}_{1}^{\prime}-\hat{N}_{1}^{\prime\prime}-(m_{1}^{\prime}+m_{1}^{\prime\prime})^{2}](A_{2}^{(2)}+2A_{3}^{(2)}+A_{4}^{(2)}-A_{1}^{(1)}-A_{2}^{(1)})]\Bigg\},
 \end{eqnarray}
\begin{eqnarray}
b_{-}(q^{2}) & = & -a_{-}(q^{2})|_{h_{V}^{\prime\prime}\to h_{T}^{\prime\prime}}+\frac{N_{c}}{16\pi^{3}}\int dx_{2}d^{2}p_{\perp}^{\prime}\frac{h_{P}^{\prime}h_{T}^{\prime\prime}}{x_{2}\hat{N}_{1}^{\prime}\hat{N}_{1}^{\prime\prime}}\Bigg\{8(m_{2}-m_{1}^{\prime})(A_{4}^{(3)}+2A_{5}^{(3)}+A_{6}^{(3)})\nonumber \\
 &  & -6m_{1}^{\prime}(A_{1}^{(1)}+A_{2}^{(1)})+4(2m_{1}^{\prime}-m_{1}^{\prime\prime}-m_{2})(A_{3}^{(2)}+A_{4}^{(2)})\nonumber \\
 &  & +2(3m_{1}^{\prime}+m_{1}^{\prime\prime}-2m_{2})(A_{2}^{(2)}+2A_{3}^{(2)}+A_{4}^{(2)})\nonumber \\
 &  & +\frac{2}{w_{V}^{\prime\prime}}\Bigg[2[M^{\prime2}+M^{\prime\prime2}-q^{2}+2(m_{1}^{\prime}-m_{2})(m_{1}^{\prime\prime}+m_{2})]\nonumber \\
 &  & \times(A_{4}^{(3)}+2A_{5}^{(3)}+A_{6}^{(3)}-A_{3}^{(2)}-A_{4}^{(2)})+2Z_{2}(3A_{4}^{(2)}-2A_{6}^{(3)}-A_{2}^{(1)})\nonumber \\
 &  & +2\frac{q\cdot P}{q^{2}}\Bigg(6A_{2}^{(1)}A_{1}^{(2)}-6A_{2}^{(1)}A_{2}^{(3)}+\frac{2}{q^{2}}(A_{1}^{(2)})^{2}-A_{1}^{(2)}\Bigg)\nonumber \\
 &  & +[q^{2}-2M^{\prime2}+\hat{N}_{1}^{\prime}-\hat{N}_{1}^{\prime\prime}-(m_{1}^{\prime}+m_{1}^{\prime\prime})^{2}+2(m_{1}^{\prime}-m_{2})^{2}] (A_{2}^{(2)}+2A_{3}^{(2)}+A_{4}^{(2)}-A_{1}^{(1)}-A_{2}^{(1)})\Bigg]\Bigg\}.
\end{eqnarray}
The
$A_{j}^{(i)}$ in the above equations are given as follows:
\begin{eqnarray}
A_{1}^{(1)} & = & \frac{x_{1}}{2},\quad A_{2}^{(1)}=A_{1}^{(1)}-\frac{p_{\perp}^{\prime}\cdot q_{\perp}}{q^{2}},
A_{1}^{(2)}  = -p_{\perp}^{\prime2}-\frac{(p_{\perp}^{\prime}\cdot q_{\perp})^{2}}{q^{2}},\quad
A_{2}^{(2)}=(A_{1}^{(1)})^{2}, A_{3}^{(2)}=A_{1}^{(1)}A_{2}^{(1)},\nonumber \\
A_{4}^{(2)} & = & (A_{2}^{(1)})^{2}-\frac{1}{q^{2}}A_{1}^{(2)},\quad A_{1}^{(3)}=A_{1}^{(1)}A_{1}^{(2)},A_{2}^{(3)}=A_{2}^{(1)}A_{1}^{(2)},A_{3}^{(3)}=A_{1}^{(1)}A_{2}^{(2)},A_{4}^{(3)}=A_{2}^{(1)}A_{2}^{(2)},A_{5}^{(3)} = A_{1}^{(1)}A_{4}^{(2)},\nonumber \\
A_{6}^{(3)} & = & A_{2}^{(1)}A_{4}^{(2)}-\frac{2}{q^{2}}A_{2}^{(1)}A_{1}^{(2)},Z_{2} =  \hat{N}_{1}^{\prime}+m_{1}^{\prime2}-m_{2}^{2}+(1-2x_{1})M^{\prime2}+(q^{2}+q\cdot P)\frac{p_{\perp}^{\prime}\cdot q_{\perp}}{q^{2}}.
\end{eqnarray}

The explicit forms of $h_{M}^{\prime}$ and $w_{M}^{\prime}$ are
given by~\cite{Cheng:2003sm}
\begin{eqnarray}
 &  & h_{P}^{\prime}=h_{V}^{\prime}=(M^{\prime2}-M_{0}^{\prime2})\sqrt{\frac{x_{1}x_{2}}{N_{c}}}\frac{1}{\sqrt{2}\tilde{M}_{0}^{\prime}}\varphi^{\prime},\nonumber \\
 &  & h_{S}^{\prime}=\sqrt{\frac{2}{3}}h_{\,^{3}A}^{\prime}=(M^{\prime2}-M_{0}^{\prime2})\sqrt{\frac{x_{1}x_{2}}{N_{c}}}\frac{1}{\sqrt{2}\tilde{M}_{0}^{\prime}}\frac{\tilde{M}_{0}^{\prime2}}{2\sqrt{3}M_{0}^{\prime}}\varphi_{p}^{\prime},\nonumber \\
 &  & h_{\,^{1}A}^{\prime}=h_{T}^{\prime}=(M^{\prime2}-M_{0}^{\prime2})\sqrt{\frac{x_{1}x_{2}}{N_{c}}}\frac{1}{\sqrt{2}\tilde{M}_{0}^{\prime}}\varphi_{p}^{\prime}\nonumber \\
 &  & w_{V}^{\prime}=M_{0}^{\prime}+m_{1}^{\prime}+m_{2},\quad w_{\,^{3}A}^{\prime}=\frac{\tilde{M}_{0}^{\prime2}}{m_{1}^{\prime}-m_{2}},\quad w_{\,^{1}A}^{\prime}=2,
\end{eqnarray}
where $\varphi^{\prime}$ and $\varphi_{p}^{\prime}$ are the light-front
momentum distribution amplitudes for $s$-wave and $p$-wave mesons,
respectively~\cite{Cheng:2003sm}:
\begin{eqnarray}
 &  & \varphi^{\prime}=\varphi^{\prime}(x_{2},p_{\perp}^{\prime})=4\left(\frac{\pi}{\beta^{\prime2}}\right)^{3/4}\sqrt{\frac{dp_{z}^{\prime}}{dx_{2}}}\exp\left(-\frac{p_{z}^{\prime2}+p_{\perp}^{\prime2}}{2\beta^{\prime2}}\right), \varphi_{p}^{\prime}=\varphi_{p}^{\prime}(x_{2},p_{\perp}^{\prime})=\sqrt{\frac{2}{\beta^{\prime2}}}\varphi^{\prime}, \frac{dp_{z}^{\prime}}{dx_{2}}=\frac{e_{1}^{\prime}e_{2}}{x_{1}x_{2}M_{0}^{\prime}}.
\end{eqnarray}

%\end{appendices}

\end{document}